# Machine learning approaches to identify thresholds in a heat-health warning system context

Pierre Masselot[1*], Fateh Chebana[2], Céline Campagna[3], Éric Lavigne[4], Taha B.M.J. Ouarda[2], Pierre Gosselin[2,3,5]

[1]Department of Public Health, Environment & Society, London School of Hygiene & Tropical Medicine, London, UK

[2]Institut National de la Recherche Scientifique, Centre Eau-Terre-Environnement, Québec, Canada;

[3]Institut National de Santé Publique du Québec, Québec, Canada;

[4]School of Epidemiology and Public Health, University of Ottawa, Ottawa, Canada;

[5]Ouranos, Montréal, Canada.

*Corresponding Author: pierre.masselot@lshtm.ac.uk

    15-17 Tavistock Pl, London WC1H 9SH, UK

# Abstract

During the last two decades, a number of countries or cities established heat-health warning systems in order to alert public health authorities when some heat indicator exceeds a predetermined threshold. Different methods were considered to establish thresholds all over the world, each with its own strengths and weaknesses. The common ground is that current methods are based on exposure-response function estimates that can fail in many situations. The present paper aims at proposing several data-driven methods to establish thresholds using historical data of health issues and environmental indicators. The proposed methods are model-based regression trees (MOB), multivariate adaptive regression splines (MARS), the patient rule-induction method (PRIM) and adaptive index models (AIM). These methods focus on finding relevant splits in indicator data but do it in different fashions. A simulation study and a real-world case study hereby compare the discussed methods. Results show that proposed methods are better at predicting adverse days than current thresholds and benchmark methods. The results nonetheless suggest that PRIM is overall the more reliable method with low variability of results according to the scenario or case.

**Keywords**: Adaptive index models (AIM) ; Multivariate adaptive regression spline (MARS) ; Patient rule-induction method (PRIM) ; Regression trees; Threshold; Warning system.

## 1. Introduction

Following several major and deadly heat wave events, such as the European one of August 2003 (*e.g.* Valleron and Boumendil, 2004; Conti *et al.*, 2005), the occurrence and impact of heat waves have become widely studied topics (Perkins and Alexander, 2013; Ouarda and Charron, 2018). Indeed, it is now accepted that heat waves are major environmental stressors that have important public health consequences, mainly in terms of excess mortality (Guo *et al.*, 2017; Mora *et al.*, 2017). In addition, the frequency and intensity of heat waves are both expected to increase with climate change (IPCC, 2013), with potentially major public health consequences (Guo *et al.*, 2018).

To adapt to the increasing frequency and in severity of heat waves, many countries or regions across the world have implemented heat-health warning systems (HHWS) (Casanueva *et al.*, 2019). The purpose of a HHWS is to anticipate potentially adverse heat waves and take appropriate action to reduce its impact on mortality or other health issues. HHWSs include mainly three components that widely vary among regions: i) the tracking and forecasting of a heat indicator, ii) one or several alert thresholds that indicate when heat is expected to a danger to the population and iii) an issuance of warning to the concerned stakeholders and targeted populations (WMO, 2015). Indicators include simple temperature measures, more sophisticated heat indices such as the Humidex (Provençal *et al.*, 2016), UTCI (Pappenberger *et al.*, 2015), or the more climate-based synoptic approaches (Sheridan and Kalkstein, 2004). Thresholds also vary based on the local population characteristics and susceptibility to heat (Gosling *et al.*, 2017). As an example, the city of Montreal (Canada) issues warnings when forecasted three days means of minimum and maximum temperature both exceeds their respective thresholds of 20 and 33 °C (Chebana *et al.*, 2013). Warning issuance can then take diverse forms and strengths, from public communication constraints, targeted calls to vulnerable populations (Mehiriz *et al.*, 2018) and hospital management (Toutant *et al.*, 2011).

Now that the HHWS have been implemented for at least 10 years in most developed countries, it appears that their effectiveness is mixed and vary from region to region (Toloo *et al.*, 2013). Some studies find slight reductions in extreme heat-related mortality (e.g. Benmarhnia *et al.*, 2016; Martínez-Solanas and Basagaña, 2019) while others struggle finding some (e.g. Weinberger *et al.*, 2018). Although communication and intervention strategies can be an explanation of these mixed results, it has been noted that alert thresholds do not always accurately correspond to temperatures at which adverse effects can be observed (Vaidyanathan *et al.*, 2019). Appropriate thresholds are indeed important for a positive impact of HHWS, and should seek a trade-off between the detection of potentially adverse heat waves, while avoiding too frequent alerts that may cause fatigue in the population (Baseman *et al.*, 2013).

The method to fix alert thresholds vary between locations, with some choosing an elevated temperature percentile (usually the 95$^{th}$ one, such as Belgium), others basing the thresholds on a few selected past events (such as France and the province of Quebec, Canada, Chebana *et al.*, 2013; Pascal *et al.*, 2006), and some on regression models within predefined weather types (the synoptic systems, Sheridan and Kalkstein, 2004). Therefore, many of the existing thresholds have been set through ad-hoc methods and with little statistical evidence. Nonetheless, research is still ongoing to propose thresholds based on historical heat-response relationships (Cheng *et al.*, 2019; Islam *et al.*, 2017; Petitti *et al.*, 2016).

The common ground of all the methods mentioned above is their two-step structure in which: i) knowledge about the association between temperature and mortality is learned, usually through epidemiological models and ii) a threshold is selected from this association using diverse criteria such as confidence intervals. The shortcomings of this structure are that the first step is not specifically focused on thresholds and thus that the uncertainty of the epidemiological analysis adds to the uncertainty of the criterion chosen to select appropriate thresholds.

The objective of the present paper is to propose methods that focus directly on threshold identifications for HHWS. The proposed methods should be data-driven to limit their subjectivity and focus directly on threshold estimation without any need for prior modelling of the relationship. In this context, the rationale guiding the choice of methods is to exploit the nonlinearity of temperature-mortality associations to directly find an appropriate level of heat above which the associated mortality sharply increases. Although this idea underlies recent work proposing thresholds (Cheng *et al.*, 2019; Longden, 2018), the proposed methods should also work well with any type of heat index, and allow the inclusion of thresholds on several variables. The above rationale leads us to regression tree-related methods that include: a) model-based recursive partitioning algorithm (MOB, Zeileis *et al.*, 2008), b) multivariate adaptive regression splines (MARS, Friedman, 1991; Weber *et al*., 2012), c) the patient rule-induction method (PRIM, Friedman and Fisher, 1999; Polonik and Wang, 2010) and d) the adaptive index model (AIM, Tian and Tibshirani, 2011).

Section 2 details the statistical problem of estimating thresholds and how to apply MOB, MARS, PRIM and AIM in this context. These methods are then compared through a simulation study in section 3. Section 4 discusses a case study on data from the metropolitan area of Montreal (Canada) to show a practical application of the four proposed methods and compare their outcome to the current thresholds. Then, section 5 discusses the results and their implication for the general context of warning systems.

## 2. Threshold estimation problem

### 2.1. The threshold model

The association between daily temperature $X$ and a health outcome of interest $Y$ is usually modelled nonlinearly as:

$$g(\mathbb{E}(Y)) = \beta_0 + f(X; \boldsymbol{\beta}) + \boldsymbol{\gamma}^T \boldsymbol{C} \qquad (1)$$

In most applications, including the one below, $Y$ represents daily count of deaths either of all cause of for specific causes, and the link function $g$ is then the $log$ function for quasi-Poisson regression that accounts for overdispersion (Bhaskaran *et al.*, 2013). However, the framework introduced here can be used for different outcome variables such as counts of hospital admissions (*e.g.* Vaidyanathan *et al.*, 2019; Yan *et al.*, 2020) or emergency calls (Islam *et al.*, 2017). The parameter $\beta_0$ is an intercept, and the vector $\boldsymbol{\beta}$ represents the parameters used to represent $f$ (usually through splines) and can also be expanded to include lags of $X$ (Gasparrini *et al.*, 2010). The vector $\boldsymbol{C}$ contains confounders such as splines of time, day-of-week, and daily humidity, with associated coefficients $\boldsymbol{\gamma}$.

Although the association between $X$ and $Y$ is usually modelled nonlinearly, it has been shown that it can be well approximated by a piecewise linear function (Cheng *et al.*, 2019; Longden, 2018; Masselot *et al.*, 2018). Thus, we consider the following model:

$$g(\mathbb{E}(Y)) = \beta_0 + \boldsymbol{\gamma}^T \boldsymbol{C} + \begin{cases} \boldsymbol{\beta_2} X & \text{when } X \geq s \\ \boldsymbol{\beta_1} X & \text{otherwise} \end{cases} \qquad (2)$$

where the link function $g$, the intercept $\beta_0$, and the vector $C$ are the same as in equation (1). In the context of an HHWS, controlling for these variables reduces the risk of predicting extremes that would not be due to heat. .

The indicators $X$ are the variables on which the decision to launch an alert is based and can be any measure of temperature, as well as any other indicators that account for specific characteristics of the location and alert system. For instance, Chebana et al. (2013) consider $X = (Tmin; Tmax)$ that are respectively daily minimum and maximum temperature in their method, and some application may also include humidity in the variables. We assume here that all variables in $X$ are either continuous or ordinal variables. As we consider machine-learning methods, we allow variables in $X$ to be correlated in order to be able to apply the methods with Tmin and Tmax for instance. Although the methods discussed allow for any number of variables in $X$, in practice the number is rarely above three. Besides, considering too much variables would increase the chances of falling in the curse-of-dimensionality pitfall.

The objective of the present study is to estimate the vector $s$ that represents the thresholds above which the risk associated to temperature variables $X$ changes. By focusing on the hot part of the year, we assume that $\boldsymbol{\beta}_2 > \boldsymbol{\beta}_1$. Note that model (2) can easily be extended to several thresholds $s_j$ ($j = 1, \ldots, J$), but we only consider the largest one with the highest risk $\beta_j$.

## 2.2. Thresholds assessment

Although we focus here on threshold determination, the end goal is to predict adverse conditions due to temperature (typically over-mortality days). When thresholds are set, it is thus important to assess them by comparing the historical days for which $X \geq s$ to the historical over-mortality days.

In the present study, agreement between predicted alerts and true extreme days are measured by sensitivity, precision, and F-score.

Sensitivity is defined as the proportion of true extreme days that are predicted by the method. The precision is the proportion of predicted alerts from a method that are true extreme days. A method with high sensitivity will thus detect most extreme days, while a method with high precision will rarely predict false alerts thus avoiding alert fatigue. The F-score measures the trade-off between sensitivity and precision and is defined as (e.g. Hripcsak and Rothschild, 2005):

$$F = \frac{2 \times sensitivity \times precision}{precision + sensitivity} \qquad (3)$$

which ranges from zero to one, with zero meaning that all alerts are false alerts and one meaning that alerts include all and only true extreme days.

## 3. Threshold determination methods

To estimate a threshold as defined in model (2), we consider methods from different statistical frameworks that all have in common an underlying objective of finding appropriate splits in the data. The simplest one may be the threshold regression, or breakpoint estimation when the variable is time, that seeks a break in a linear relationship. However, this framework tend to consider each variable separately, which is why we instead consider the related framework of regression trees through the MOB algorithm (Zeileis *et al.*, 2008) that iteratively splits the region spanned by ***X*** in order to account for interaction.

An alternative way of looking at splits, is to consider them as knots in a linear spline context. We thus consider the MARS algorithm that iteratively adds knots to linear splines while also accounting for interaction between variables in $X$. Finally, we also consider the framework of rule estimation, where the rules are of the form $X \geq s$. The PRIM algorithm seeks to find such rules in order to maximize an objective function, while the AIM algorithm uses such rules for prediction.

### 3.1. Model-based recursive partitioning (MOB)

Regression trees represent a popular body of machine-learning methods, whose idea is to partition the data along the exposure dimension and fit a simple model inside each group (Loh, 2014). The splits created by such an algorithm are thus good candidate for thresholds since they represent a break in the exposure-response function. There are many popular regression trees algorithm (such as CART, CTREE or GUIDE, see Loh, 2014, for an extensive literature review on the subject) but we hereby consider the MOB algorithm of Zeileis *et al*. (2008) that directly uses information from a fitted model to partition the data and thus is adapted to the expectation that the impact of heat accelerate above the threshold.

The algorithm starts by considering the whole dataset as a single group which is split in two groups along one variable in $X$. Each of the two groups can in turn be split along either the same variable or another variable in $X$, and so on resulting in a tree structure. To perform the splitting, MOB fits a simple parametric model at each node and performs a change-point test (Zeileis *et al.*, 2003) on the residuals of the fitted model versus each of the variables in $X$. If the test is significant on at least one variable of $X$ the node should be split and the splitting variable is the one with the smallest p-value on the test. The location of the split is then determined through a line search in order to minimize the model error in the two children nodes. The process is then repeated iteratively until none of the change-point tests are significant.

In the context of a HHWS, at each node we consider a quasi-Poisson linear model as expressed in model (2) to grow the tree, i.e. with variables in $X$ and $C$ as the predictors. In typical applications, the variables in $X$ are also those along which the splitting is performed, although it can be slightly different according to the context with, e.g., different temperature indicators used as predictor and splitting variables. We consider the typical significance level of 5% and, as MOB tend to construct much smaller trees than CART for instance, we do not perform any pruning after the tree is grown. We restrict the splits such that each node contains at least 5 observations to allow extreme thresholds with at least a few occurrences.

Once the tree is grown, the selected thresholds are the extreme splits on the path leading to the node in which the risk $\beta_2$ is the highest. Since some variables in $X$ may never be split on the path leading to this extreme node, MOB allows variable selection for thresholds. Thus, in the present context we consider that if a variable has no selected threshold in the end, it should be dropped from the HHWS. In the present study, MOB is fit with function glmtree in the R package partykit (Hothorn and Zeileis, 2015), with default parameters unless specified otherwise.

### 3.2. Multivariate adaptive regression splines (MARS)

The MARS algorithm (Friedman, 1991) is a trade-off between regression trees and splines expansion. The algorithm seeks to fit splines of the form $(X_j - t_j)_+$ and $(t_j - X_j)_+$ to high dimensional data, by selecting only a low number of knots $t$. The algorithm thus searches the optimal $t_j$ to approximate the relationship between $Y$ and the predictor variables $X_j$ in a forward stepwise manner. It starts with an empty model and recursively adds a knot on each of the predictor variables in $X$ to the model. At each step, the chosen variable and the knot are chosen such that it results in the most important error decrease of the final model. It thus grows a large model of $k$ terms of the form $Y = \beta_0 + \sum_k \beta_k B_k(X)$ in which the $B_k(X)$ are spline bases. The $\beta_k$s are estimated at each step using a simple linear regression. To account for interactions between variables in $X$, products between spline basis functions of different indicators are also explored for addition to the model. At the end, some terms are discarded by a backward deletion to keep only the most predictive bases, usually a small number.

The attractiveness of such a procedure for HHWS is that the chosen knots are natural candidates as alert thresholds and that it also accounts for interaction between the temperature measures in $T$, the indicators in $X$ and the covariates in $C$. In the context of HHWS, we estimate the quasi-Poisson model (1) by MARS by allowing interactions up to degree 2. The thresholds are then the extreme knots found by MARS for each of the variables in $X$. Since each split created in the forward pass of the algorithm concerns the whole domain of variables, and not a specific subset as in regression trees, the most extreme splits can end up creating a group without any observations. In this case, we select lower splits on one or several variables of $X$ in order to maximize the average of $Y$ while the number of observations exceeding all thresholds is above five. As for MOB, a lack of split in a variable $X_j$ is considered as a discard of this variable that thus should not be used in the HHWS. In the present study, MARS is fit with R package earth (Milborrow, 2018) with default parameters unless specified otherwise.

### 3.3. Patient rule-induction method (PRIM)

PRIM is a bump-hunting algorithm that seeks to find boxes in the indicator space $X$ inside which an objective function of the response $f(Y)$ is high (Friedman and Fisher, 1999). The boxes are defined as upper and lower limits for each variable in $X$, i.e. $X_j \in [t_{j1}; t_{j2}]$ to allow straightforward interpretation of the results. The PRIM algorithm starts with a box containing the whole dataset and iteratively peels the box such that the objective function $f(Y)$ increases inside it. At each step of the peeling, a small portion $\alpha$ of the data is removed either to the left or to the right of one variable. The variable and the side of the peeling are chosen in order to obtain the peeled box with the highest value of $f(Y)$. The algorithm then continues until the number of observations remaining inside the box is below a predetermined number. At the end, a pasting process is usually carried out. It consists in refining the edges of the box by slightly expanding it as long as it increases $f(Y)$ inside the box.

In the context of an HHWS, instead of the mean, a quasi-Poisson linear model as in (2) is fitted on the observations contained in the current box. Thus, at each iteration, the model is peeled in order to maximize the risk $\beta_2$. Such a criterion is proposed by LeBlanc *et al.* (2005) in the context of survival analysis. Since we are interested in heat thresholds, we consider only left peeling, which means that only the lower bound of the box is peeled. The process is carried out until the proportion of observations left in the box is below $\phi_0$. The final box is then chosen as the one yielding the largest rate of increase compared to the previous larger box as (Dazard *et al.*, 2016):

$$\frac{\beta_2^k - \beta_2^{k-1}}{\phi_{k-1} - \phi_k} \tag{4}$$

where $\beta_2^k$ is the risk and $\phi_k$ the proportion of remaining observations in the box at step $k$ of the peeling. The largest rate of increase (4) thus represents the best change-point in the relationship between $T$ and $Y$. Once this box is chosen, the final thresholds are the lowest bound of the box for each of the variables in $\boldsymbol{X}$.

In the present study, we consider $\alpha = 5\%$ as it has been shown to yield good performances (Chong and Jun, 2008; Abu-Hanna *et al.*, 2010; Capurso *et al.*, 2016) and $\phi_0 = 5/n$ with $n$ the total number of observation. This allows the algorithm to consider boxes with as low as 5 heat wave days, while the optimal number is chosen by the criterion (4). The method is provided in a custom R package primr available online (https://github.com/PierreMasselot/primr).

### 3.4. Adaptive index models (AIM)

The objective of AIM (Tian and Tibshirani, 2011) is to derive indices in the form of simple binary rules that best predict the response of interest with the model:

$$g(\mathbb{E}(Y)) = \beta_0 + \beta_1 \sum_k \mathbf{1}(X_{j(k)} \geq t_k) \tag{5}$$

where $\mathbf{1}(.)$ is the index function that take 1 when its argument is true, $j(k)$ is the variable chosen at index $k$, $t_k$ is split point, while $\beta_0$ and $\beta_1$ are the usual regression coefficients. Note that the AIM allows the same variable to enter the model several time. To construct model (5), at each step the indicator variable $X_{j(k)}$ and $s_k$ are chosen in order to maximize the statistic testing the hypothesis $\beta_1 = 0$, i.e. to maximize the fit of the AIM to the response $Y$ (Huang *et al.*, 2017). The algorithm thus iteratively adds new terms until a maximum number of terms $K$ is reached. In practice, $K$ is chosen through cross-validation.

For a HHWS, we fit the AIM of Equation (5) using all the variable in Equation (2) to account for covariates in $C$. The thresholds $s$ are then the extreme split point associated to the variables in $X$. In the present study we apply the AIM with gaussian response since no method for Poisson response is available. Note however, that in many practical cases we take advantage of the convergence of the Poisson distribution to a gaussian one when the rate increases. We allow up to three splits per variable $X_j$ in the model, with the total number of splits chosen by 5-fold cross-validation. The model is fit with the R package AIM (Tibshirani, 2010).

## 4. Simulation study

In the present section, a simulation study is conducted to assess the strengths and weaknesses of each considered method for estimating a threshold.

### 4.1. Design

The design of the simulation study attempts to emulate real data in environmental epidemiology. We generate $p$ ($p = 1,2,3,4$) indicators $X_1, \ldots, X_p$ that all have a basic linear association with a response $Y$, with an additional impact when all indicators cross predefined thresholds $s_1, \ldots, s_p$. It can be expressed as the following data-generating mechanism:

$$Y = \sum_{j=1}^{p} \beta_{1j} X_j + \beta_2 (\boldsymbol{X} - \boldsymbol{s})_+ + \epsilon \tag{6}$$

where $(\boldsymbol{X} - \boldsymbol{s})_+ = (\boldsymbol{X} - \boldsymbol{s}) I(\boldsymbol{X} \geq \boldsymbol{s})$, with $I(.)$ the indicator function, is the joint impact of the indicators' extreme values and $\epsilon$ is a Gaussian white noise.

Indicators $\boldsymbol{X}$ are generated from a multivariate normal distribution with potentially non-null covariance between the variables. We fix the value of $\beta_{1j} = 1$. When all variables in $\boldsymbol{X}$ exceed their respective thresholds, the risk increases by $\beta_2 \geq 0$. It represents an additional impact of extremes. The thresholds $\boldsymbol{s}$ are set such that around 1.5% of observations fall in the extreme category.

We consider variations of three parameters in the general data generating mechanism of Equation 6: a) the number $p$ of variables in $\boldsymbol{X}$, with $p = 1, \ldots, 4$, b) the magnitude $\beta_2$ of the extreme impact with $\beta_2 = 0.1, 0.2, 0.3, 0.5, 1$, and c) the correlation $\rho$ between variables in $\boldsymbol{X}$ with $\rho = 0, 0.5$. The two latter parameters control the impact of extreme indicators values on the response $Y$. A full factorial design of these parameters thus results in 40 different scenarios.

For each scenario, a large number $B = 1000$ datasets of length $n = 1000$ are generated. Each method is then applied to all datasets, resulting in $B = 1000$ estimated thresholds $s_b$ for each method and each scenario. We then compute the alerts detected by these thresholds to compare them with the true extreme days generated through the criteria given in section 2.2. As benchmark methods, we consider the segmented regression of Muggeo (2003) and the GAM-based criterion of Petitti et al. (2016). The segmented regression is a very flexible method among the large family of threshold regression. It focuses on the location estimation of one or several breakpoints in a regression line and allows several variables as well as a non-Gaussian response, in our case Poisson. We automatically estimate the number of thresholds, by starting with 10 thresholds for each variable in $X$ and discarding non-admissible breakpoints at each iteration of the fitting process, as described by Muggeo and Adelfio (2011). The final thresholds are then the most extreme ones among the breakpoints fitted by the algorithm.

To represent thresholds estimated through a complete exposure-response fitting, we apply a GAM with Poisson response between $Y$ and $X$. The thresholds are then computed as the lowest value above the minimum mortality temperature such that the lower confidence interval of the estimate exposure-response function is above the zero-line (Petitti *et al.*, 2016). This procedure is hereby selected as it represents a common intuition to determine HHWS thresholds (e.g. Islam *et al.*, 2017; Vaidyanathan *et al.*, 2019).

### 4.2. Results

Figure 1 shows the mean and 95 % interval of F-score for each method and each scenario. Overall, the most stable and consistent method is PRIM that shows the highest F-scores when $p \geq 2$, closely followed by MARS. They both present consistently higher F-scores than the two benchmark methods that are SEG and GAM. Figures S1 and S2 shows that PRIM and MARS provide good trade-offs between sensitivity and precision.

On the other hand, the MOB algorithm performances are heavily dependent on the magnitude of extreme effect with lowest F-scores for a 10% increase above the thresholds but the highest scores at a 100% increase. Figures S1 and S2 show that MOB tend to have high sensitivity but low precision, which is probably due to the stringency of its splitting criterion which tends to often not select some variables. The AIM algorithm overall presents the lowest F-scores of the four proposed methods with lower performances than SEG.

Figure 1 indicates that the F-score is at its highest for $p = 1$ and decreases as $p$ increases. The score is also slightly higher when there is correlation between the variables in $X$, as shown by the bottom row of panels in Figure 1. Among the discussed methods, PRIM is the one with the lesser drop in performances when $p$ increases. The only exception to the overall rule is GAM that shows the highest F-scores for $p = 4$ and no correlation. However, this is the less likely scenario in practice.

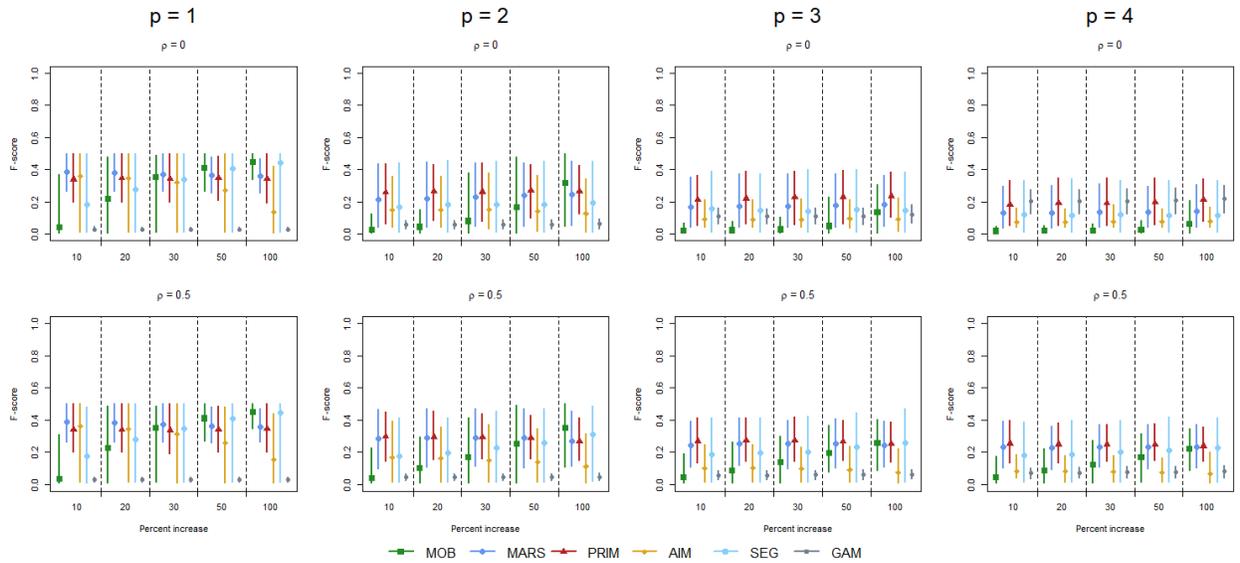

**Figure 1: Mean and 95% confidence interval of F-scores versus the percentage increase of risk above the threshold for each simulation scenario. $p$: number of variables in $X$; $\rho$: correlation between variables in $X$.**

## 5. Application to Montréal's HHWS

In this section, the proposed methods are applied to determine objective thresholds for an HHWS in the city of Montreal, Canada, and compared to the current thresholds.

### 5.1. Data and application

We consider daily data ranging from May to September of years 1990 to 2014 of several administrative health regions around the city of Montreal, Canada. The health regions incorporated are *Montreal*, *Laval*, *Lanaudière*, *Montérégie* and the south of the *Laurentides* region; these administrative regions constitute the metropolitan region of Montreal. They correspond to class 1 in Giroux *et al*. (2017) and represented a large population of 3 209 173 in 1990 and 4 015 900 in 2014. This gathering of regions allows to benefit from a large number of cases inside a small area in which the weather can be considered almost uniform, and to compare the computed thresholds to the ones currently used in the area. An alert is.

The health outcome corresponding to $Y$ in Equations (1) and (2) is the daily count of all cause of deaths in the Montreal region. The temperature variables in $X$ are weighted moving average of daily minimum and maximum temperatures (respectively referred to as Tmin and Tmax in the following). Both Tmin and Tmax use the same weights that are (0.4,0.4,0.2) for lags 0 to 2 (Chebana *et al.*, 2013). These indicators are those used in the current HHWS, with an alert launched when the Tmin and Tmax indicators exceed 20 and 33°C respectively, which will correspond the benchmark thresholds for comparison purposes. Note that in MOB and PRIM, the quasi-Poisson linear models estimating $\boldsymbol{\beta}_1$ and $\boldsymbol{\beta}_2$ are fitted with the single temperature measure $T$ as $Tmean = (Tmin + Tmax)/2$ instead of both Tmin and Tmax because of their high collinearity (Barnett *et al.*, 2010). This stabilizes the algorithms while being roughly equivalent. The covariates considered in $C$ are two natural spline components: one on the day of season (i.e. 1 for the 1st of May, 2 for the 2nd of May, etc.) with 4 degrees of freedom and one on the year with one degree of freedom per decade (as in, e.g. Gasparrini, *et al.*, 2015). These splines allow accounting for seasonal patterns and long-term trend in mortality. As the underlying population changes only slightly from day to day and that its trend is captured by the spline components of time in $C$, no offset is considered here which is common in time series study in environmental epidemiology (Bhaskaran *et al.*, 2013).

Since the purpose of an HHWS is to prevent excesses of mortality from the baseline, we compare the methods on their ability to detect accurately historical over-mortality (OM) events. It is computed as $OM = 100 \times (Y - EM)/EM$ (Chebana *et al.*, 2013), where $EM$ is a baseline of expected mortality computed using spline smoothing, with the same bases as the covariates considered above. It thus allows for an event detected in the middle of summer where the base mortality is lower to be as important as at the edges of the summer season. The OM events are defined by different cut points from 30 to 50, and the ability of each method to detect these events is measured by the sensitivity, precision and F-score as explained in section 2.2. The thresholds and scores obtained by the considered methods are compared to those of the currently implemented thresholds in the region of Montreal (thereafter referred to as *Reference*). These thresholds were obtained through the method of Chebana et al. (2013) that is based on a prior determination of over-mortality level and a line search of thresholds in order to maximize both sensitivity and specificity criteria (see section 2.2).

The variability of each method for both the estimated thresholds and alert prediction accuracy is assessed through bootstrap simulations. In this analysis, $B = 1000$ bootstrap replications are generated from the present data and the methods are applied on each replication to obtain a set of bootstrap thresholds. Since the present data are time series, we use a non-overlapping block bootstrap (Carlstein, 1986) with a block being a year of data. This means that years of data are resampled with replacement to create the bootstrap replications, which allows keeping the time-dependent structure of data.

## 5.2. Results of the proposed methods

Applying the MOB algorithm to the Montreal data leads to the tree shown in Figure 2. Here, the alert system is represented by the extreme terminal node defined by the single splitting rule $Tmax \geq 32.7°C$. It contains 11 observations and the scatterplot indicates that this is the node with the steepest risk associated to Tmean. Note that the fitted tree shows well the initial assumption of piecewise relationship with increasing risk above each split, since the slope is slightly decreasing for the lowest temperatures (defined by $Tmin \leq 4.5 \ °C$) and increases from the leftmost panel to the rightmost one. Since the extreme terminal node does not involve $Tmin$ in its splitting sequence, we consider it as a variable selection indicating that $Tmin$ is not needed in the HHWS. However, if a threshold on $Tmin$ is needed, its best split as a surrogate to the one reported here would be $Tmin \geq 17.5$.

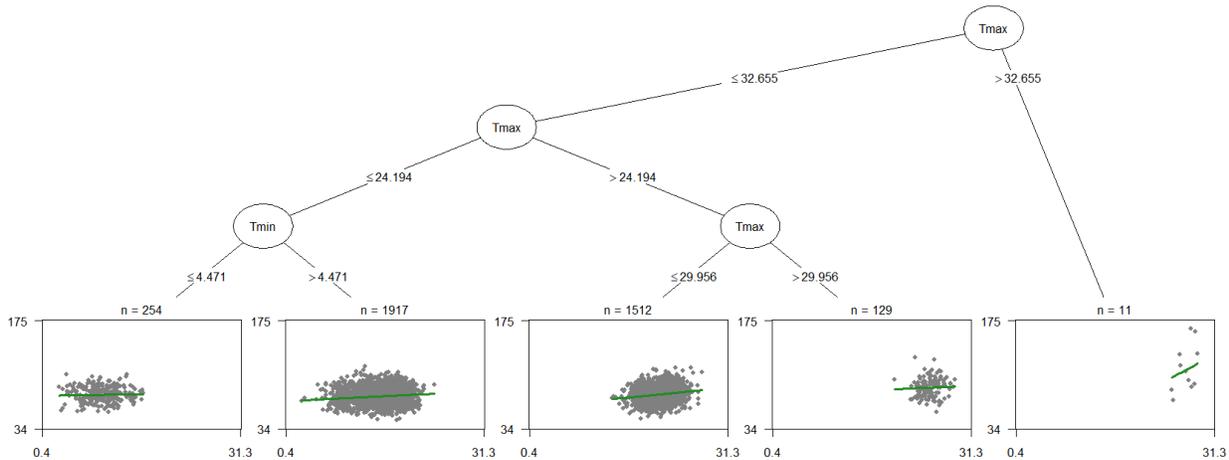

**Figure 2: Tree grown by the MOB method. Terminal leaves shows $Y$ against Tmean for the observation in each node, with the fitted regression line for Tmean in green.**

Figure 3 shows the exposure-response function induced by MARS for the two variables in $X$. Overall the surface increases with both $Tmin$ and $Tmax$ although it is steeper and with more knots along the $Tmax$ dimension, in particular at extreme values. The alert region thus corresponds to the upper right corner above the thresholds of 18.4°C and 32.7°C of the surface which contains 10 observations.

Figure 4 shows the end of the peeling trajectory estimated by PRIM. This trajectory shows two main breakpoints. The first one is just above the 0.5 % support (i.e. 22 remaining observations) and the second one is the highest at a support of around 0.3% that corresponds to 11 remaining observations. The latter yields thresholds of 20.3°C and 32.1 °C. Note that this trajectory could easily be used to define additional alert level.

Figure 5 shows the score surface of the AIM. Here, we consider the portion with the highest scores in the top right corner. This identifies thresholds at 20.4°C and 32.2°C respectively.

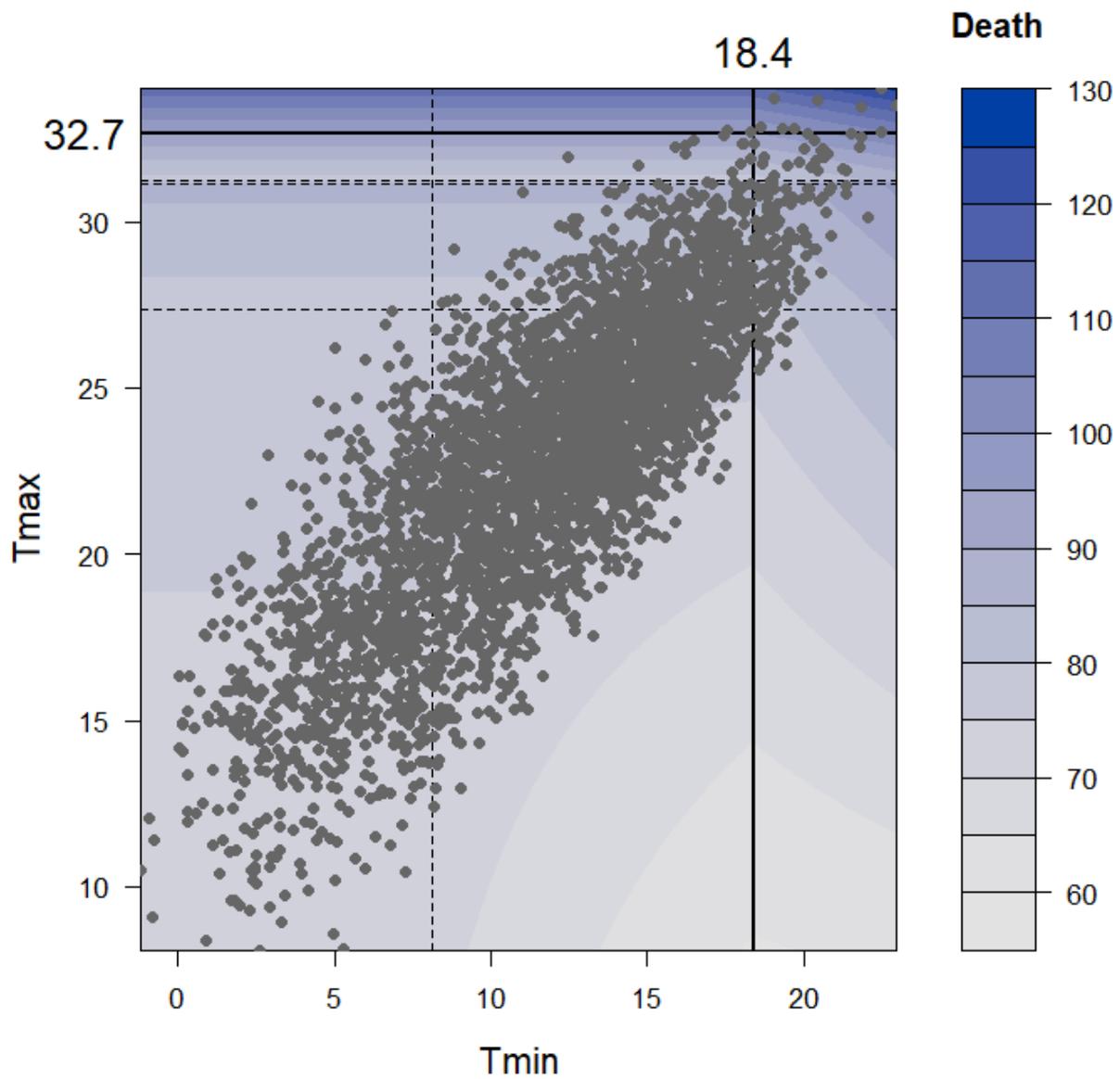

**Figure 3: Exposure response surface fitted by MARS for the indicators $Tmin$ and $Tmax$. Dashed lines indicates knots selected by the algorithm with thick continuous lines indicating extreme knots kept as thresholds.**

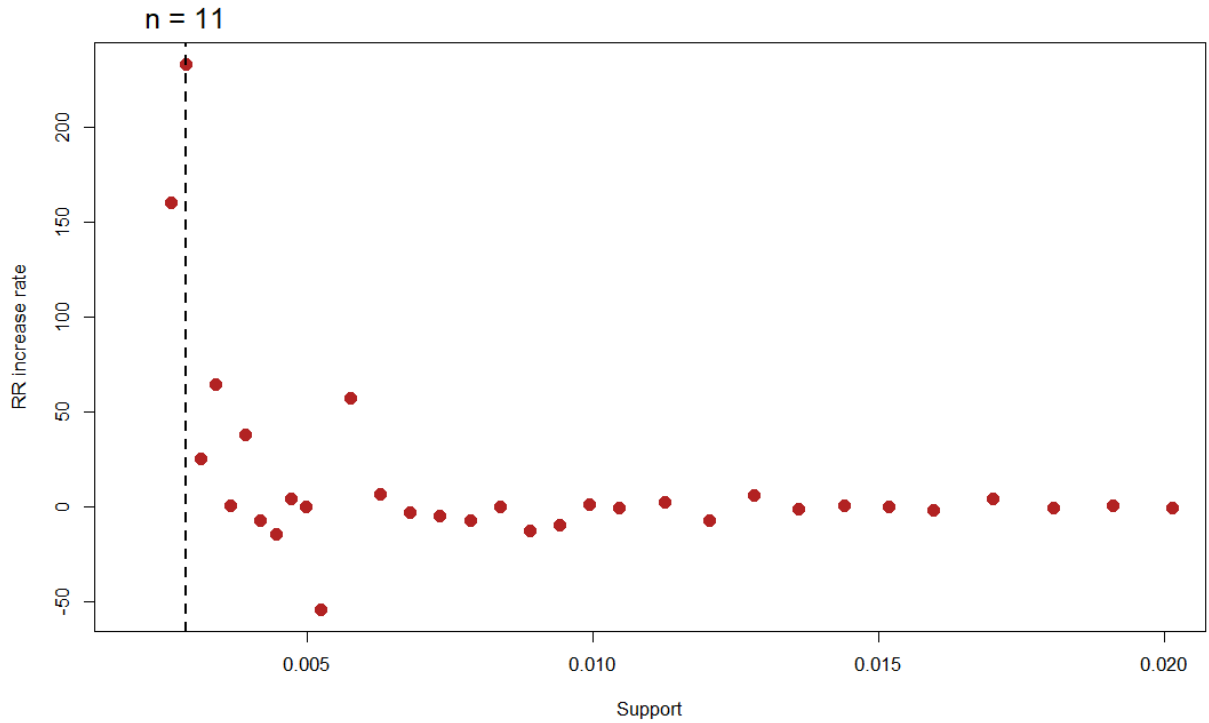

**Figure 4: Peeling trajectory of PRIM. The ordinates show the increase rate of RR associated to *Tmean* compared to the previous box in the peeling process.**

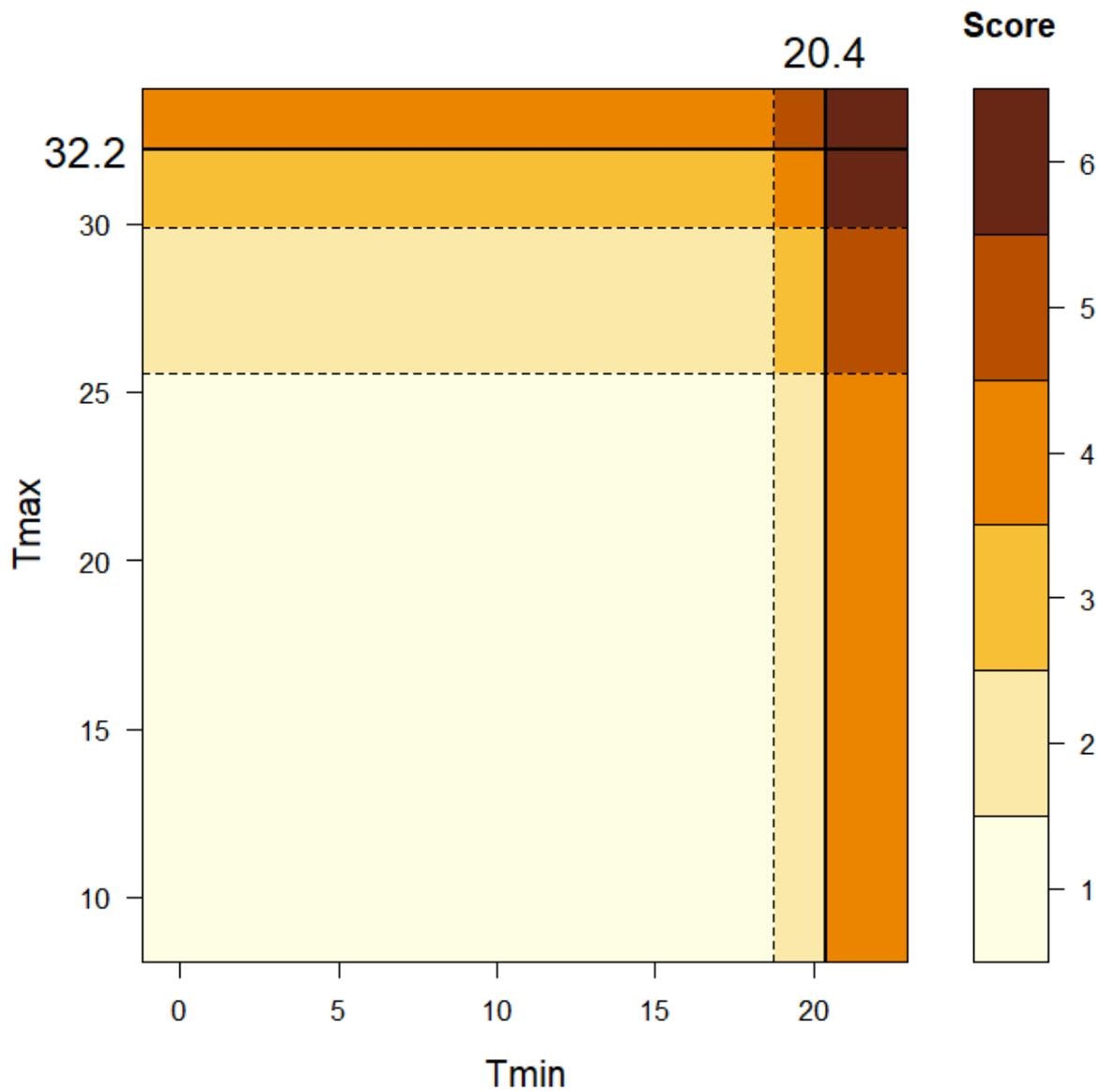

**Figure 5: AIM scores for each value of Tmin and Tmax with day-of-season and year kept constant at their middle point. Lines indicate the cutpoints locations with thick continuous ones the final thresholds.**

**Table 1 : Estimated thresholds, alerts and performances for the proposed methods applied to the HHWS of Montreal. 95% bootstrap confidence intervals are in parenthesis. * Reference are thresholds and performances resulting from the work of Chebana *et al.* (2013). ** Number of days corresponds to the absolute number of exceedances while episodes correspond to gathering of exceedance days, separated by less than 3 days. *** OM: Over-mortality**

| | Thresholds | | Alerts** | | | |
|---|---|---|---|---|---|---|
| Method | Tmin | Tmax | Number of days | Number of episodes | Mean OM (%)*** | Coverage (%) |
| MOB | - | 32.7 (29.2-32.7) | 12 (10-126) | 9 (6-63) | 42.6 (12.4-58.4) | 13.4 (6.2-45.2) |
| MARS | 18.4 (4.7-21.8) | 32.7 (24.4-32.6) | 10 (10-103) | 7 (5-49) | 47.0 (7.4-55.1) | 12.3 (5.3-37.9) |
| PRIM | 20.3 (19.3-21.8) | 32.1 (29.2-32.6) | 11 (8-25) | 5 (4-16) | 43.9 (8.6-59.3) | 12.6 (3.1-22.0) |
| AIM | 20.4 (18.4-20.4) | 32.2 (28.2-32.3) | 9 (12-275) | 5 (6-97) | 49.7 (8.5-59.5) | 11.7 (13.3-69.8) |
| Reference* | 20.0 | 33.0 | 4 | 3 | 73.3 | 7.7 |

Table 1 summarizes the thresholds found by each method with the resulting number of historical alerts as well as the mean over-mortality of these alert days. MOB and MARS find the same threshold for Tmax which is the highest one of all proposed methods. However, the MARS algorithm also estimates a threshold on Tmin, resulting in a more stringent alert system and thus less historical episodes compared to MOB. The PRIM and AIM algorithms propose lower thresholds on Tmax but higher ones in Tmin than MARS and MOB. The result is a similar number of alert days, but gathered in less episodes than MARS and MOB (only 5 here). The ranking of average OM of alert days detected by the four proposed methods is inversely proportional to the number of detected alert days. It ranges from 42.6% for MOB to 49.7% for AIM, which means that the extra days detected by MOB likely have lower OM, but not by much.

Table 1 also reports coverage values in order to measure the accuracy of predicted alerts. For a method $m$, the coverage is defined as $OM_m * n_m/n$ where $OM_m$ and $n_m$ are the mean OM and number of alerts for the method while $n$ is the total number of observations. Coverage measures the trade-off between a high number of detected alerts and significant enough alerts, the higher the coverage, the higher average OM compared to the number of alert detected. In spite of overall lower OM average, MOB results in the highest coverage, followed by PRIM, which means that the extra alert detected by the two methods are relevant for an HHWS. On the other hand, AIM presents the lowest coverage value with a difference of only 1.7% below the best method (MOB). The Reference coverage is even below the one of AIM, because of the low number of alert days.

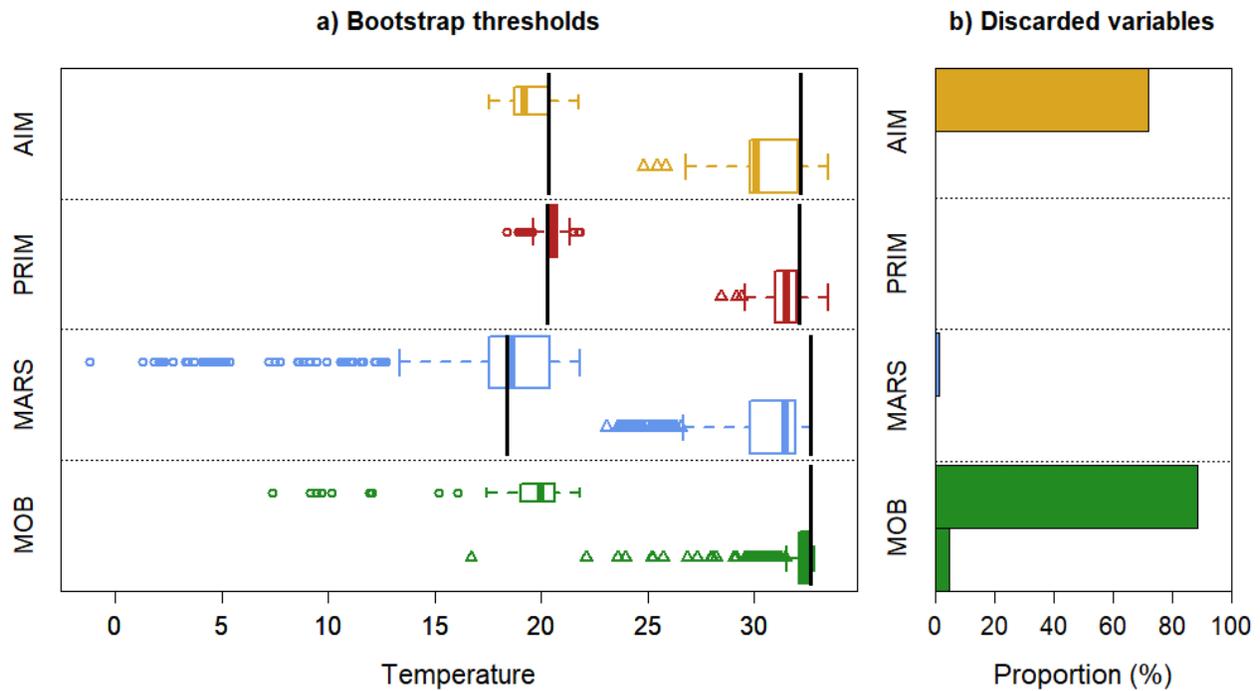

**Figure 6: Boxplots of the bootstrapped thresholds (a) and proportion of replications in which the method discards each variable (b). In both cases, the upper values represent Tmin and the lower represent Tmax. Vertical solid lines indicate the threshold found on the whole sample and reported in Table 1.**

## 5.3. Performance comparison

Boxplots of Figure 6a show the distribution of the $B = 1000$ bootstrap thresholds and Figure 6b the proportion of samples in which the variable was not selected by the model. Overall, PRIM seems to be the least biased method, as the observed thresholds are close to the mean of bootstrap replications. In addition, the threshold distributions of PRIM show little outliers meaning that PRIM has also a low variance. MOB also shows little bias for the Tmax threshold with some outliers, and very few Tmin thresholds. Finally, MARS and AIM seem to be the most biased and variable methods overall.

Figure 7 shows the three considered criteria for several cut points $u$ ranging from 30 to 50 %, for each method. For low values of $u$, MOB shows the highest sensitivity and is equal to AIM and PRIM for the highest values of $u$. On the contrary, the Reference show the lowest sensitivity values, meaning that these thresholds miss important OM events. The Reference shows however the highest precision due to the low number of alert days. The F-score values clearly indicate that for low cut points $u$, MOB is the best method, while for high cut points, its F-scores are lower than the other considered methods. At high cut points values, AIM present the highest F-scores.

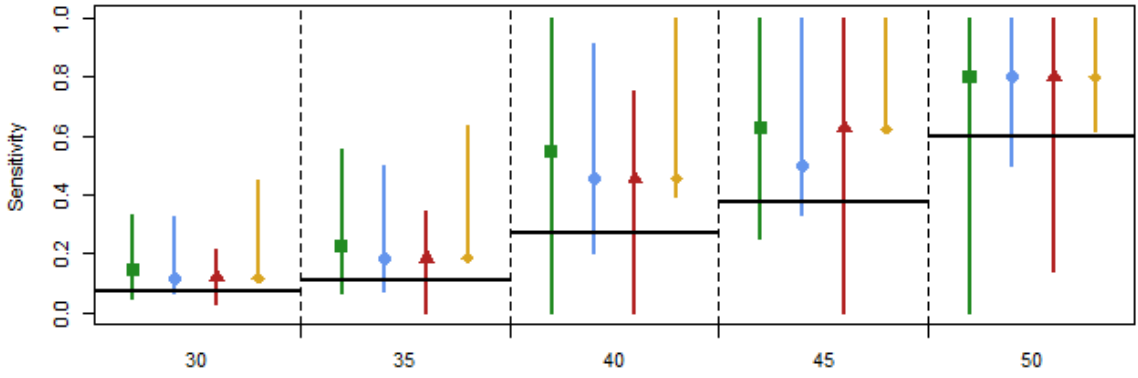

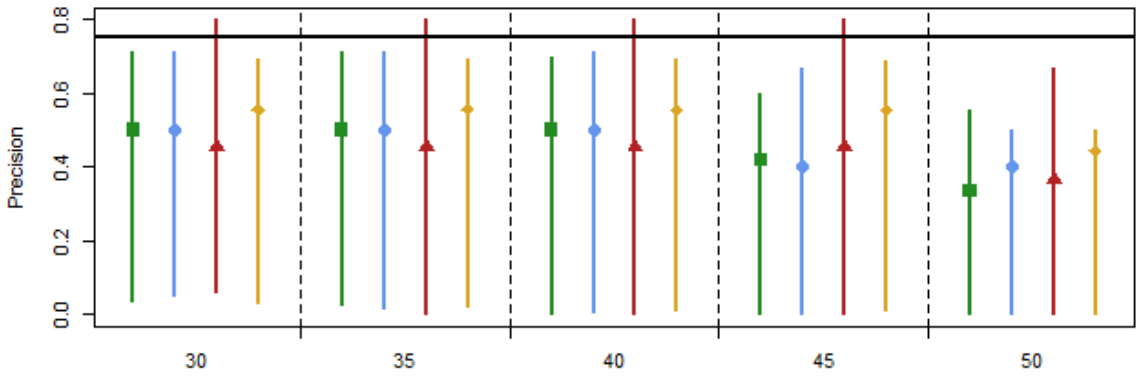

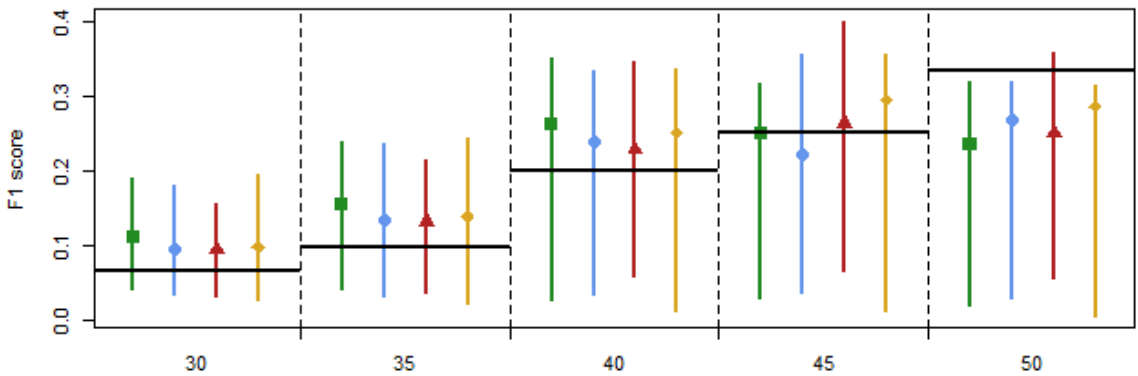

**Figure 7**: Whole sample value and 95% interval from bootstrap replications of:. a) sensitivity which is the proportion of OM peaks detected by the system, b) precision which is the proportion of alerts that are actual OM peaks and c) $F_1$-score making a trade-off between the two. The x-axis represents the OM cut-point above which it is considered an over-mortality event, For each measure, the closer to 1 the value, the better the thresholds.

## 6. Discussion

The present paper considers four different methods for data-based threshold identification: MOB, MARS, PRIM and AIM. The four methods are applied to a large number of simulated datasets and to a health data case study, and compared using a number of criteria. The results of both the simulation study and the case study example show that the four methods can result in appropriate thresholds with very little prior choices. In comparison, methods based on prior models such as GAM may not always yield a clear enough relationship to derive threshold from it, an issue concerning rural areas in particular because of lower counts.

Comparisons between the methods show that the overall best method is PRIM, presenting the highest prediction power in simulations, and a good fitting of observed high OM events in the case study. In addition, the method is the most robust to design changes and varies little on bootstrap resampling, while offering important flexibility to the user with the peeling trajectory or other criteria such as a cross-validated trajectory (Dazard *et al.*, 2016). MOB shows the best performances on the case study while having a good prediction power in specific scenarios of the simulation study. By opposition to PRIM which guarantees a threshold is found, MOB allows threshold selection that can be useful in exploratory analyses to determine relevant heat indicators. The latter may be adapted for subgroup identification in clinical studies where there are a lot of biomarkers to choose from, but perhaps less in environmental epidemiology where the number of indicators is lower and each indicator could be of interest. Note however that, although not explored here, the MOB algorithm also allows exploring breakpoints through the statistic used to choose each split in the tree. Such a tool may allow for greater flexibility as well. Finally, MARS and AIM show lower performances by being slightly less adapted to the specific objective of threshold determination. In particular, by considering the creation of thresholds independently from other variables, it can result in a set thresholds with no observation exceeding all of them, especially when there are more than two variables in *X*.

In practice, it may be useful to obtain uncertainty measures of the estimated thresholds such as standard deviations or confidence intervals. Here we tackled this matter through bootstrap replications, but uncertainty could also be measured through bayesian versions of the discussed methods (excluding AIM, Chipman *et al.*, 1998; Denison *et al.*, 1998; Wu and Chipman, 2003). On the other hand bagging is a possibility to obtain more robust thresholds based on bootstrap replications (Breiman, 1996). Note that bagging regression trees basically corresponds to the so-called random forests, although the well-known algorithm is based on CART (Breiman, 2001) instead of MOB. Alternatively, boosting the algorithms could also improve the robustness of estimated threshold (Friedman, 2001; Wang *et al.*, 2004). However, this exceeds the scope of the present paper.

The present work does not address the matter of temporal dependence in the data that can create clusters of alerts. Future work should nonetheless address the subject of temporal dependence. For instance, criteria based on episodes rather than on days considered as independents could be used. Modelling the response as a Markov chain, in a similar fashion as in the extreme value theory literature (Winter and Tawn, 2016) is also a possible lead.

The present study focuses on unidimensional outcomes, mainly mortality although all the methods proposed are applicable to other typical health outcomes such as morbidity and emergency calls.

However, it can also be of interest to determine thresholds based simultaneously on several of these outcomes and as with finer causes accounting for multi-morbidity. Since PRIM focuses on an objective function of the response, it can easily be extended to multiple outcomes. On the other hand, the task is less trivial for other algorithms, which is beyond the scope of the present paper.

Although the present paper focuses on the heat wave issue, the possibility of tuning algorithms for better flexibility of the proposed methods allows their application to other issues. For instance, there is a growing interest in developing public health action plans and thresholds related to cold spells (Bustinza and Lebel, 2013; Yan *et al.*, 2020). Thresholds for waterborne diseases and extreme precipitations (*e.g.* Guzman Herrador *et al.*, 2015), hip fractures and snow conditions (Modarres *et al.*, 2014) as well as air pollution and respiratory diseases (De Sario *et al.*, 2013; Masselot *et al.*, 2019) are also of interest. These thresholds could be based on actual regional health impacts and exposure instead of a general national standard with inherent high variability.

The present paper focuses on threshold estimation, so other aspects of a methodology to set HHWS have not been thoroughly discussed. An example is the creation of indicators for $X$, but this aspect is beyond the scope of the present paper. The present study considered the same indicators (Tmin and Tmax) as in Chebana *et al*. (2013) but other possibilities could arise. For instance, different lags of the indicators can be included directly in the methods, and scores can then be used to weight these lags and construct indicators with associated thresholds. Such scores include variable importance for regression trees (Ishwaran, 2007) and variable relevance for PRIM (Friedman and Fisher, 1999).

## Acknowledgements

The authors would like to thank the Ouranos consortium for its financial and scientific partnership in this study, as well as the Quebec government Climate change Action Plan 2013-2020 financial contribution. The authors also want to thank the *Institut national de santé publique du Québec* for access to mortality data. The authors also acknowledge the important role of Jean-Xavier Giroux and Christian Filteau for extracting the data and Anas Koubaa for helping in obtaining preliminary results. Finally, the authors warmly thank Joint Editor Jouni Kouha and two anonymous reviewers for their help in improving the manuscript.

All analyses are performed with the open source software R (R Core Team, 2020) with the addition of the packages `partykit` for applying MOB (Hothorn and Zeileis, 2015), `earth` for applying MARS (Milborrow, 2018), `AIM` for applying AIM (Tibshirani, 2010) and custom code for applying PRIM, available on the corresponding author's github (https://github.com/PierreMasselot/primr). The codes written for both the simulation study and case study are also freely available on the first author's github (https://github.com/PierreMasselot/Paper--Machine-Learning-Thresholds).

Sheridan, S. C. and Kalkstein, L. S. (2004) Progress in Heat Watch–Warning System Technology. *Bull. Am. Meteorol. Soc.*, **85**, 1931–1941. DOI: 10.1175/bams-85-12-1931.

Tian, L. and Tibshirani, R. (2011) Adaptive index models for marker-based risk stratification. *Biostatistics*, **12**, 68–86. DOI: 10.1093/biostatistics/kxq047.

Tibshirani, L. T. and R. (2010) *AIM: AIM: Adaptive Index Model*. Available at: https://CRAN.R-project.org/package=AIM (accessed 13 November 2019).

Toloo, G., FitzGerald, G., Aitken, P., et al. (2013) Evaluating the effectiveness of heat warning systems: systematic review of epidemiological evidence. *Int. J. Public Health*, **58**, 667–681. DOI: 10.1007/s00038-013-0465-2.

Toutant, S., Gosselin, P., Bélanger, D., et al. (2011) An open source web application for the surveillance and prevention of the impacts on public health of extreme meteorological events: the SUPREME system. *Int. J. Health Geogr.*, **10**, 39. DOI: 10.1186/1476-072x-10-39.

Vaidyanathan, A., Saha, S., Vicedo-Cabrera, A. M., et al. (2019) Assessment of extreme heat and hospitalizations to inform early warning systems. *Proc. Natl. Acad. Sci.*, **116**, 5420–5427. DOI: 10.1073/pnas.1806393116.

Valleron, A.-J. and Boumendil, A. (2004) Épidémiologie et canicules : analyses de la vague de chaleur 2003 en France. *C. R. Biol.*, **327**, 1125–1141. DOI: 10.1016/j.crvi.2004.09.009.

Wang, P., Kim, Y., Pollack, J., et al. (2004) Boosted PRIM with application to searching for oncogenic pathway of lung cancer. In: *Proceedings. 2004 IEEE Computational Systems Bioinformatics Conference, 2004. CSB 2004.*, August 2004, pp. 604–609. DOI: 10.1109/CSB.2004.1332514.

Weber, G.-W., Batmaz, I., Koksal, G., et al. (2012) CMARS: a new contribution to nonparametric regression with multivariate adaptive regression splines supported by continuous optimization. *Inverse Probl. Sci. Eng.*, **20**, 371–400. DOI: 10.1080/17415977.2011.624770.

Weinberger, K. R., Zanobetti, A., Schwartz, J., et al. (2018) Effectiveness of National Weather Service heat alerts in preventing mortality in 20 US cities. *Environ. Int.*, **116**, 30–38. DOI: 10.1016/j.envint.2018.03.028.

Winter, H. C. and Tawn, J. A. (2016) Modelling heatwaves in central France: A case-study in extremal dependence. *J. R. Stat. Soc. Ser. C Appl. Stat.*, **65**, 345–365. DOI: 10.1111/rssc.12121.

WMO (2015) *Heat waves and health: guidance on warning-system development*. World Meteorological Organization;World Health Organization.

Wu, L. and Chipman, H. (2003) *Bayesian Model-Assisted PRIM Algorithm*. Technical Report. Waterloo, Ontario, CA: University of Waterloo.
36